# Pain pathogenesis in rheumatoid arthritis – what have we learned from animal models

Emerson Krock, Alexandra Jurczak, Camilla I Svensson

Department of Physiology and Pharmacology, Centre for Molecular Medicine, Karolinska Institutet, Stockholm, Sweden

**Corresponding author**: Prof. Camilla I. Svensson

Department of Physiology and Pharmacology

Center for Molecular Medicine, L8

Karolinska Institutet, Stockholm, Sweden

Phone: +46 8 524 87948

Email: camilla.svensson@ki.se

**Conflicts of Interest**: None

**Abstract**

Rheumatoid arthritis (RA) is a systemic autoimmune disease characterized by joint inflammation and joint pain. Much of RA treatment is focused on suppressing inflammation, with the idea being that if inflammation is controlled other symptoms, such as pain, will disappear. However, pain is the most common complaint of RA patients, is often still present following the resolution of inflammation, and can develop prior to the onset of inflammation. Thus, further research is needed to better understand RA-associated pain mechanisms. A number of preclinical rodent models commonly used in rheumatology research have been developed based on bedside-to-bench and reverse translational approaches. These models include collagen-induced arthritis, antigen-induced arthritis, streptococcal cell wall-induced arthritis, collagen antibody-induced arthritis, serum transfer from K/BxN transgenic mice and tumor necrosis factor (TNF)-transgene mice. They have led to increased understanding of RA pathogenesis and have aided the development of successful RA treatments. More recently, these models have been used to elucidate the complexities of RA associated pain. Several potentially modifiable mechanisms, other than inflammation, have been investigated in these models and may in turn lead to more effective treatments. Furthermore, preclinical research can indicate when specific treatment strategies may benefit specific patient subgroups or at which disease stage they are best used. This review not only highlights RA-associated pain mechanisms, but also suggests the usefulness of preclinical animal research based on a bedside-to-bench approach.

# 1. Introduction

Rheumatoid arthritis (RA) is a debilitating systemic disease characterized by joint inflammation (edema and synovitis), fatigue and joint pain. If left untreated, the disease leads to bone and cartilage destruction and increased mortality. In developed countries RA affects 0.5-1% of the population and is approximately 3 times more common in women than men. Joint pain, termed arthralgia, commonly develops prior to disease onset [66] and can change in quality and intensity during the course of the day and as the disease progresses [22,26,65,70]. The introduction of new disease-modifying antirheumatic drugs (DMARDS) has dramatically improved the prognosis of RA patients. More problematic however is that a significant proportion of patients fail to report long-term suppression of pain that reflects the degree of disease control. Even if treatment with DMARDs reduces pain during the course of treatment, a proportion of RA patients are dissatisfied with their pain management and continue to rate pain relief as one of their top priorities for improved health and quality of life [6,97,164,170].

The etiology of RA is unclear, but genetic and environmental factors are thought to drive autoimmune-directed pathological events. Antibodies that recognize endogenous proteins (i.e. autoantibodies) in the sera is an underlying characteristic of ~70% of RA cases, and autoantibodies can be present for years prior to RA diagnosis [141]. Seropositive RA is associated with an increased risk for bone erosion and joint damage [67], while seronegative RA patients often present with higher inflammatory activity [130]. Rheumatoid factor (RF) directed against the Fc domain of IgG and anti-citrullinated protein antibodies (ACPA) that bind proteins where arginine has been converted to citrulline (e.g. vimentin, fibrinogen, alpha-enolase) are the most prominent autoantibodies in RA and are used for diagnosis. In addition, autoantibodies against cartilage components, such as anti-collagen type II (CII) antibodies, and autoantibodies against post-translational modifications other than citrullination, such as anti-carbamylated protein antibodies (anti-CarP) have been described in RA patients [40,69,82,123,155,164].

Autoantibodies exert their actions by directly binding their antigens (i.e. collagen), which can result in functional modifications of the antigen. They also form immune complexes (IC, a complex of antibodies bound to an antigen) that activate the complement system and Fc receptors (FcRs). Interestingly, autoantibodies may be coupled to pathology and pain in the absence of overt joint inflammation. Arthralgia and autoantibodies, especially ACPA, are strong predictors of RA development and this period before diagnosis is termed 'pre-RA' [18,106]. Transfer of ACPA to naive mice induce both pain-like behaviour and bone erosion without causing synovitis [92,180]. ACPA can however prolong LPS-induced inflammation and pain-like behaviour, or worsen existing synovitis [94]. Thus, multiple "hits" may be necessary for the development of RA. One potential scenario is that the presence of certain autoantibodies at the time of a trauma or infection lead to a non-resolving immune response and persistent pain.

The inflammatory events in RA involve both innate and adaptive immunity. The innate immune response includes macrophages, and dendritic cells, which produce a cascade of inflammatory mediators that attract neutrophils and cells of the adaptive immune system, T and B cells, to the site of injury. The environment shaped by this process will ultimately lead to the destruction of the joint. Central to RA pathogenesis is the release of proinflammatory cytokines, such as tumor necrosis factor (TNF) and interleukins 1, 6, and 17 (IL-1, IL-6, and IL-17). Therapies targeting inflammatory processes including conventional DMARDS like methotrexate, and biologic DMARDs, such as cytokine blocking anti-TNF therapies (i.e. etanercept and infliximab) and IL-6 signaling inhibition (i.e. tocilizumab), are often successful at controlling joint inflammation (reducing swelling and synovitis) and limiting joint destruction [176,179]. However, as mentioned, subpopulations of patients continue to report pain as problematic [97,98,170]. An increasing amount of evidence indicates that RA-associated pain is not solely driven by peripheral inflammatory processes. A number of recent studies suggest central pain processing is altered in RA [55,99]. For example, people with RA display reductions in pressure and thermal pain-thresholds and increased pain sensitivity over inflamed joints, as well as at remote, non-articular sites [53,139]. Neuropathic pain symptoms have only been sporadically tested in RA patients but studies using the painDETECT questionnaire show that a subpopulation of RA patients may have a neuropathic component [5,28,56,90,144]. Up to 20% of RA patients fulfill the criteria of "likely neuropathic pain" and direct measurements revealed that 33% of people with RA who reported neuropathic symptoms displayed clinical evidence of neuropathy [157]. The fact that autoantibodies, inflammation, and pain do not always go hand in hand indicates that RA pain mechanisms are more complex than simply being driven by inflammation. Thus, advancing our understanding of the relation between RA pathology and persistent joint pain is critical for development of improved pain relief strategies.

Over the past 40 years, a number of arthritis models have been developed in the rheumatology field based on common characteristics of RA such as autoimmunity, synovitis, and bone and cartilage degradation [13,174]. These models have contributed to the development of several biologic DMARDs that are used today [113], such as anti-TNF therapies [84,118,181]. However, when therapies were being developed, joint inflammation was the primary consideration and pain was rarely evaluated. When evaluating arthritis-associated pain mechanisms, models of inflammatory pain induced by intra-articular injections of Complete Freund's Adjuvant (CFA) or carrageenan are frequently used. However, these models are based on local activation of the innate immune system and do not include aspects of an adaptive immune response. Thus intra-articular injection of CFA or carrageenan may not be the best strategy to investigate RA-associated pain mechanisms. The use of different models in RA research and pain research has in some ways led to a disconnect between fields. One may argue that it is the local innate inflammation that alters neuronal excitability in arthritis and the way of generating such a state is not critical. Of course, no single animal model will recapitulate all aspects of human pathology but using a number of clinically informed models likely has a greater chance to include important aspects of the disease mechanisms that are critical for "chronification" of pain in arthritis. Thus, a reverse translational research (i.e. bedside-to-bench) approach using animal models that are based on and informed by the symptoms, mechanisms and characteristics of RA identified in patients is of value to identify therapeutic strategies to manage pain. [1,46]. The following is a brief introduction to RA models that have been utilized in pain studies (Table 1). We then highlight pain-related findings from these models with an emphasis on cytokines, receptors and ion channels.

## 2. Animal Models of RA

### 2.1 Immunization Models

Collagen type II (CII) is one of the main extracellular matrix proteins in articular cartilage and therefore anti-CII antibodies target the joint. Autoantibodies against CII are found in serum and synovial fluid of a subset RA patients [40,69,82,123], including at diagnosis [40,107,123]. Collagen induced arthritis (CIA) is the most frequently used experimental model of RA [126]. It is established by immunization with CII in complete CFA which leads to production of anti-CII antibodies directed to the same major CII epitopes as in RA patients [101]. In mice and rats, polyarthritis develops within 3-4 weeks after immunization and many important features of human RA are recapitulated, including T- and B-cell activation and mononuclear cell infiltration into synovium, joint cavity and periarticular structures. Synovial oedema, hyperplasia, and cartilage and bone erosion also occur in CIA [44,145]. CIA mice display decreased climbing, locomotion and grooming behaviours [73], thermal hypersensitivity [73,83], and mechanical hypersensitivity, which develops prior to inflammation [11,12] and lasts for at least 28 days following arthritis-onset [73].

Antigen-induced arthritis (AIA) and streptococcal cell wall (SCW) induced arthritis are T-cell mediated monoarthritis models commonly used in mice and rats [20,21,78,86,152]. The rodents are immunized against the antigen (methylated bovine serum albumin/ovalbumin and SCW, respectively). In AIA the antigen is mixed with CFA and inoculated at the base of the tail. The antigen is then injected alone into the joint three weeks later. In SCW-induced monoarthritis SCW fragments are injected intra-articularly followed by low dose intravenous (i.v.) challenges of SCW ("reactivation") giving rise to flair ups. Both models are characterized by a transition from acute phase with joint swelling and immune cell infiltration to a chronic stage with synovial hyperplasia, cartilage loss and bone erosion in the sensitized joint. The joint inflammation is caused by local antigen-specific T-cell hyperreactivity and also deposition of immune complex (formations of antibodies bound to the antigen) in AIA[12,41]. While pain-like behaviour has been assessed extensively in the AIA model, there are only few reports on SCW-arthritis induced pain-like behaviour. Nevertheless, in the AIA model mechanical hypersensitivity (up to 8 weeks), thermal hypersensitivity (6 weeks), and altered gait and guarding behaviour have been observed in the paw of the inflamed knee joint [15,189]. In the SCW-arthritis model mechanical hypersensitivity follows the timing of inflammatory flares [25].

**2.2 Transfer Models**

The presence of anti-CII antibodies in RA patients and mice subjected to CIA and the association between these and joint pathology have formed the basis for the collagen antibody-induced arthritis (CAIA) model [126,163,169]. A cocktail of anti-CII antibodies (IgG2a and IgG2b) is injected i.v. or intraperitoneally (i.p.) and three to five days later lipopolysaccharide (LPS) is injected i.p. to synchronize the onset of joint inflammation and increase the incidence and severity of the arthritis [126]. The K/BxN model is another transfer model of arthritis that is based on transfer of serum instead of antibodies. K/BxN transgenic mice spontaneously develop RA-like pathology [27,91]. Transfer of serum from K/BxN mice to naive recipient mice via one or two i.p. injections induces reproducible arthritis [75,105]. Anti-glucose-6-phosphate isomerase (GPI) antibodies (IgG1) together with inflammatory factors in the serum drives development of the arthritis [27,111]. Both the CAIA and K/BXN models develop swelling in small joints of the front and hind paws and sometimes in large joints within the first week after induction, with faster onset in the K/BxN model compared to CAIA. The degree of pannus formation and erosion of cartilage and bone is dependent on mouse strain and sex. Visual signs of joint inflammation resolve 2-4 weeks after induction and there is residual bone erosion detectable for several weeks after resolution. Of note, while both models mimic the effector phase of an adaptive immune response, T and B-cells *per se* only play a minor role [126]. In both models development of mechanical hypersensitivity precedes inflammation, is maintained during inflammation and persists for several weeks after the resolution of joint inflammation [3,10,17,19,29,31,165]. In addition, mice display thermal hypersensitivity [10,11,29,30] and reduced locomotor activity [19,150,158,166]

Although not yet an established model of RA, transfer of ACPA IgG isolated from patients or patient B-cell derived murinized monoclonal ACPA to naïve mice leads to mechanical and thermal hypersensitivity, decreased locomotor activity and bone erosion in the absence of joint inflammation or sickness [92,180]. The pain-like behavior develops within 3-7 days and lasts up to 28 days post injection. Thus, transfer of ACPA can be used as a model to investigate "pre-RA" arthralgia and how autoantibodies induce joint pain independent of inflammation.

**2.3 Genetic models**

Several strains of transgenic mice overexpressing single or multiple copies of a 3'-modified human TNF transgene (TNF-tg) have been developed to investigate the role of TNF in RA pathology [80,100]. At 3-4 weeks of age these mice spontaneously develop progressive inflammatory arthritis with ankle joint swelling, synovial hyperplasia and immune cell infiltration evident and by 9-10 weeks pannus formation, bone and cartilage destruction and fibrosis detectable. The TNF-tg mice display mechanical and thermal hypersensitivity by week 6, which is still present at week 10. Tail-flick responses increase and motor and mobility are decreased at week 10 [68].

**3. Sex in RA Models**

Preclinical pain research has been criticized for primarily using male rodents despite female humans being disproportionality affected by chronic pain. In 2005 Mogil and Chanda found 79% of rodent studies published in PAIN over the previous ten years used males only, 8% female only, 9% male and female and in 3% sex was unspecified [120]. Mogil recently stated that these percentages were similar in 2015 [119]. In fact, recent rodent discoveries have highlighted sex difference in pain-like behaviour [108,159,160] and immunity mechanisms [108,121]. In 2014 the NIH created a policy requiring preclinical research to use male and female rodents [38]. Interestingly, Janine Clayton, the director of the US National Institutes of Health (NIH) Office of Research on Women's Health stated that research falling under the National Institute of Arthritis and Musculoskeletal and Skin Diseases, such as RA, display a better balance in terms of sex in preclinical studies [37]. This prompted us to examine the distribution of rodent sex for studies focused on pain-like behaviour in arthritis. We identified 51 studies using the models described in this review. In nearly 50% of the studies females or male and female rodents were used (Table 2). Thus, sex was better balanced compared to the average in the pain field described in 2005. Still, nearly 50% of the studies were conducted in male rodents only. Therefore, more female and female/male comparisons in arthritis-pain research are still needed, especially considering that RA disproportionately affects females.

**4. Mechanisms**

Here we review RA associated pain mechanisms by focusing on specific factors (i.e. cytokines) that have been investigated using the above models (Table 3). Comprehensive reviews of mechanisms focusing on particular cell types (i.e. glia) [11] and on RA inflammation and joint destruction can be found elsewhere [13,88,113].

**4.1 Cytokines and Chemokines**

*4.1.1 TNF*

TNF blockade reduces pain-like behavior in TNF-tg mice [68] and in male CIA, CAIA, K/BxN [11,14,29,73] and female CAIA, SCW and AIA rodents [11,16,25]. Despite minimal effect on joint inflammation, TNF inhibition during the inflammatory phase of AIA and CAIA rapidly reverses pain-like behavior and the effect is maintained in the late "non- inflammatory" stage [11,14]. Similarly, etanercept, a soluble TNF decoy receptor, attenuates mechanical hypersensitivity in the K/BxN and SCW models, but only during joint inflammation [25,29]. Traditionally the pronociceptive actions of TNF have been linked to stimulation of immune cells, chondrocytes and synoviocytes leading to production of pain-inducing factors. However, the analgesic effect of TNF inhibitors is evident prior to reduction of inflammation. Preventing activation of the neuronally expressed TNF receptors TNFRI [151,156] and possibly TNFRII [14,151] reduces neuronal excitability [73], suggesting that TNF contributes to pain signal transmission both directly and indirectly. Interestingly, RA patients treated with the TNF antibody infliximab experience pain relief prior to anti-inflammatory effects. These effects are associated with a reduction in painful stimuli-induced BOLD signal in pain-perception areas of the brain [68]. TNF-tg mice also have altered pain-related brain connectivity and activity, which is normalized by infliximab [68]. Thus, peripheral TNF blockade alters pain-related CNS activity in RA. In addition, intrathecal injection of etanercept rapidly decreases mechanical hypersensitivity [14] and spinal hyperexcitability in male AIA rats [89], suggesting that TNF also regulates pain signaling at the level of the spinal cord. Taken together, these data suggest TNF contributes to RA-associated pain by directly activating sensory neurons, through inflammatory processes and through spinal TNF signaling.

*4.1.2 IL-1β*

The IL-1 receptor (IL-1R) antagonist anakinra reduces disease activity in a relatively small subpopulation of RA patients [112]. Despite an increase of IL-1R expression in DRGs from female rats subjected to AIA [38] and periarticular tissues from male CIA rats [149], and the demonstrated direct effect of IL-1β directly on sensory neurons [43,52,109], anakinra has a limited effect on pain-like behavior. While it decreases thermal hypersensitivity, it does not reduce mechanical hypersensitivity in AIA [52,153]. Noteworthy, IL-1β levels are elevated in the cerebrospinal fluid (CSF) of RA patients [7,96] and female CIA rats [129], and IL-1β gene expression is elevated in male K/BxN spinal cords [96]. Spinal P2X7 antagonism reduces IL-1β levels, microgliosis and mechanical hypersensitivity [129] in CIA female rats suggesting that spinal IL-1β may be microglia dependent. Intrathecal injection of IL-1β increases dorsal horn neuronal activity and mechanical hypersensitivity in rats further supporting a direct role of spinal IL-1β in pain signaling [142]. Thus, these results not only explain the limited efficacy of IL-1β inhibition on RA-associated pain, but also suggest targeting central IL-1β signaling, rather than peripheral or systemic IL-1β, may be a more efficacious strategy for pain management.

*4.1.3 Il-6*

Like in RA patients, IL-6 is elevated in the synovial fluid of male CIA rats [133]. Intra-articular injection, and to a lesser extent systemic injection, of IL-6 inhibitors decrease pain-like behavior in female AIA rats, without affecting the joint inflammation [127]. Similar to TNF, this suggests a direct action of IL-6 on sensory neurons. Interestingly, IL-6 receptors are expressed by DRG neurons and satellite glial cells, both of which respond to IL-6 stimulation [49,58]. The role of IL-6 actions on satellite glial cells has not been investigated in detail, but sensory neuron-specific deletion of gp130 (IL-6 signaling transducer) attenuates pain-like behavior and joint inflammation in male and female AIA mice [51]. The role of central IL-6 in models of RA-induced pain has not yet been investigated but evidence to date indicates that peripheral IL-6 directly affects sensory neurons to drive RA-associated pain.

*4.1.4 IL-17*

T-helper 17 cells release the proinflammatory cytokines of the IL-17 family (IL-17A-F), which are involved in numerous immune regulatory functions as well as chondrocyte metabolism and osteoclastogenesis [9]. IL-17 levels are increased in the joints of male AIA mice and may contribute to arthritis pain-like behaviour via neutrophil recruitment and increasing pronociceptive mediators like TNF and IL-6 [114,143]. Indeed, IL-17 inhibition decreases neutrophil infiltration into the joint and mechanical hypersensitivity [137]. However, IL-17 can also act directly on nociceptors as DRG neurons express the IL-17 receptor A-E [50]. Experiments showing that IL-17 directly increases activity of DRG neurons via the IL-17A receptor [50,143] supporting this notion. Furthermore, male and female IL-17A knockout mice are protected from AIA-driven mechanical hypersensitivity, but not joint inflammation [50]. A central role for IL-17 signaling in models of RA-associated pain has not been investigated yet but spinal IL-17A gene expression is elevated in a neuropathic pain model in male rats [167] and IL-17 increases astrocyte activity *in vitro* [167,188]. Taken together, there are several possible peripheral mechanisms by which IL-17 can contribute to RA-induced pain and it will be exciting to see the outcome of IL-17-targeting clinical trials.

**4.1.5 IL-22**

IL-22 is produced by activated NK and T-cells and is involved in the coordination of both the adaptive and innate immune system responses. IL-22 expression is increased in the joints of male AIA mice and antibody blockade of IL-22 and IL-22 deletion protects mice from AIA-induced mechanical hypersensitivity. IL-22$^{-/-}$ AIA mice also show decreased synovitis compared to wild-type AIA mice. [138]. The mechanism by which IL-22 contributes to pain is poorly characterized and warrants for further studies.

*4.1.6 CXCL1, 2, 5 and 8*

IL-8 (CXCL8) and the murine analogues CXCL1, 2 and 5 are chemokines classically described as neutrophil recruitment and activation factors. Along these lines, CXCL1 and 5 levels are elevated in the knee and ankles of AIA mice [39,63] and drive neutrophil migration [39,63]. In turn, inhibition of CXCR1 and 2, the receptors for CXCL1, 2, 5 and 8, reduces mechanical hypersensitivity in male AIA mice [39]. Uncoupled from the neutrophil action, ACPA increases CXCL1 and CXCL2 secretion from primary osteoclast cultures. As ACPA also increases the expression of these chemokines in mouse ankle joints [92,180] and CXCR1/2 inhibition decreases both ACPA induced pain-like behaviour and bone loss in male mice [92,180] in the absence of other signs of inflammation, CXCL1 and CXCL2 are likely to contribute to RA-associated pain via multiple mechanisms. In support of this notion, CXCL1 activation of neuronal CXCR2 increases nociceptor excitability and sensitization [48,177,183,187]. A central role of CXCL1, 2, 5 or 8 in RA-associated pain signaling has not been investigated, but CXCL1 is increased in astrocytes and CXCR2 is increased in dorsal horn neurons of male mice subjected to nerve injury [190], intraplantar CFA injection [24] or bone cancer pain models [184]. Additionally, blocking spinal CXCL1 signaling decreases pain-like behavior in these models. Thus, studies targeting spinal CXCL1 signaling are warranted in RA models of pain. Current evidence indicates peripheral

CXCL1, 2, 5, and 8 drive RA associated pain via indirect mechanisms like neutrophil recruitment and painful bone loss, and by acting directly on nociceptors.

### 4.1.7 CX3CL1/Fractalkine

CX3CL1, known also as fractalkine (FKN), with its receptor CX3CR1 being expressed predominantly on microglia, is an important mediator of pain-associated neuron-glia communication in rodents [34,102,122,175,186]. Membrane bound FKN is released by the action of the metalloproteases ADAM10/17 and cathepsin S (CatS) released by endothelial cells and microglia, respectively [36,71,132]. Spinal injection of FKN gives rise to mechanical and thermal hypersensitivity and the mechanism behind this effect includes CX3CR1 activation, which triggers intracellular phosphorylation of p38 MAPK in the microglia, and stimulates the release of pro-nociceptive cytokines [33,35,117,191]. The role of CatS/FKN signaling in chronic pain behavior has been well studied in the CIA model. Both early and established mechanical hypersensitivity in CIA are strongly associated with CatS/FKN activity in reactive microglia. CatS inhibitor prevents development of pain-like behavior and spinal cord hyperexcitability when administered centrally early after immunization, without having any effects on joint inflammation [129]. Additionally, continuous i.t treatment with FKN-neutralizing antibody attenuated established hypersensitivity concomitantly with dampening the microglia response in the spinal cord [32]. This demonstrates that CX3CL1/ CX3CLR1 signaling in the spinal cord is important in establishing and maintaining pain-like behavior in this experimental model of arthritis.

### 4.2 Toll-like Receptors

Toll-like receptors (TLR) are activated by endogenous damage-associated molecular patterns (DAMPs) such as fragmented extracellular matrix components, HMGB1 and heat-shock proteins. Many DAMPs are increased in RA models [31,79,87,116] and TLRs are expressed on a number of cell types, including immune cells, glia, chondrocytes, synoviocytes, osteoclasts and sensory neurons [54,79,95]. TLR signaling increases nociceptive factors, such TNF and nerve growth factor [79,93], and joint inflammation in RA models [79,124]. In male K/BxN mice, spinal gene expression of the TLR4 ligands tenascin-c and HSP90 are increased [31] and extranuclear HMGB1 increases in male and female CAIA mice [3]. The elevated levels of DAMPS in the spinal cord appears to be linked to pain signal transmission as mechanical hypersensitivity is reduced in male TLR4$^{-/-}$ K/BxN mice and spinal inhibition of TLR4 signaling during the inflammatory phase but not the late phase of K/BxN arthritis reduces mechanical hypersensitivity [31]. Furthermore, blocking spinal HMGB1-TLR4 signaling in male and female CAIA mice reverses mechanical hypersensitivity during the inflammatory and late phases of the model [4]. Taken together, these results suggest TLR signaling mediates joint inflammation and peripheral pain through indirect, and possibly direct, mechanisms.

**4.3 Fc-gamma receptors (FcγR)**

FcγRs are widely expressed by immune cells and they bind the Fc domain of antibodies (IgG) and are activated by immune complexes (IC) in order to regulate adaptive immunity. Interestingly, the FcγRI (CD64) is expressed in a subpopulation of DRG neurons and co-expressed with nociceptive markers including TRPV1, IB4, CGRP and substance P [140]. IgG-IC stimulation of cultured DRG neurons increases Ca$^{2+}$ signaling and excitability [76,140] and intradermal injection of IgG-IC also increases *in vivo* C-fiber excitability and generates mechanical and thermal hypersensitivity [76]. In female AIA rats neuronal FcγRI expression is increased and *in vitro* and *in vivo* C-fiber excitability is increased in response to IgG-IC compared to naïve mice [76]. Furthermore, AIA rats injected with IgG-IC display increased pain-like behavior compared to AIA rats alone [76]. DRG neurons in immunized male mice also sequester and then release antigen specific IgGs that are produced by B-cells [64]. Thus, FcγRs and autoantibodies in IC formation provide a new link between the adaptive immune system and activation of sensory fibers innervating the joint that may play important roles in modulation of RA-associated pain.

**4.4 Ion Channels**

*4.4.1 Calcium Channels*

Alpha-2 delta-1 (α2δ1), an auxiliary subunit of voltage-gated $Ca^{2+}$ channels, is upregulated in DRG of neuropathic, but not usually inflammatory, pain models [11]. However, α2δ1 protein levels are elevated in DRG neurons of male CAIA mice, [165]. Gabapentin and pregabalin bind to α2δ subunits (1 and 2) and have been found to reduce calcium currents via an effect on trafficking of voltage-dependent calcium channels in the CNS [45]. Gabapentin and pregabalin also seem to be effective analgesic drugs in models of neuropathic pain, but also in some models of inflammatory pain [104,125,135,162]. Additionally, gabapentin reverses mechanical hypersensitivity in both the inflammatory and late phases of male K/BxN [29] and CAIA [10] mice. These findings demonstrate an importance of the α2δ1 subunit in arthritis-induced pain behaviour. Additionally, inhibition of store operated $Ca^{2+}$ channels with YM-58483 prevents and reverses CIA-induced pain-like behavior in male mice [57]. In a combination of the AIA and CIA models, nociceptor expression of MVIIA, which blocks the voltage gated calcium 2.2 channel, reduces abnormal weight bearing, but also increased joint destruction [8]. Taken together, these findings indicate that modulation of calcium channel activity is a potential target for arthritis pain management.

### *4.4.2 TRPV1*

TRPV1 is a nonselective cation channel expressed by nociceptors that is activated by protons, heat and endogenous lipids. It has been investigated as a target for inflammatory pain management including arthritis [161]. Pretreating animals with resiniferatoxin in order to desensitize TRPV1 and destroy nerve terminals prior to K/BxN serum transfer in male and female mice decreases mechanical hypersensitivity. Neutralization of OxPAPC, an endogenous oxidized phospholipid recently identified as a TRPV1 and TRPA1 agonist, reduces mechanical and thermal hypersensitivity in male CIA rats [131]. However, TRPV1 inactivation increases protease activity, joint swelling and bone erosion [17], which is potentially coupled to TRPV1 expression on synoviocytes [85,172], osteoclasts [147] and osteoblast-like cells [2]. These findings suggest several potentially opposing roles of TRPV1 in RA, thus inhibition could have unintended consequences.

*4.4.3 Acid Sensing Ion Channels*

Acid sensing ion channels (ASIC) are activated by decreased pH and endogenous lipids [110]. AISC3 is expressed by sensory fibers innervating the joint [72], synoviocytes [61] and osteoclasts [74]. Interestingly, ASIC3$^{-/-}$ mice are protected from CAIA-induced pain-like behavior during the inflammatory phase, but paw inflammation, bone erosion and joint destruction increase when compared to wild-type mice with CAIA [158]. While neuronal ASIC3 is linked to pain generation, it may have a protective regulatory role in synoviocytes [61] or osteoclasts.

**4.5 Nerve Injury and Neuropathic Pain Markers**

ATF3 and GAP43 are commonly used as markers of nerve injury. ATF3 expression in DRG neurons is elevated in male mice subjected to CIA [73], K/BxN [29] and CAIA [165]. GAP43 is elevated in DRG neurons during the inflammatory phase and late phase of the CAIA model [165] and GAP43 skin fiber density, a sign of peripheral neuropathy, is increased in female AIA rats, even in the absence on inflammation [77]. These findings, along with α2δ1 findings discussed above, suggest that inflammatory processes in conjunction with bone and/or cartilage remodeling in the joint introduce changes in the local environment that induces stress or is damaging to sensory neurons.

**5. Future Directions and Conclusions**

While chronic pain is an area of intense study, the translation of discoveries made in animal models into novel pain therapies for humans has been challenging. The long-term consequences of inflammatory processes in the joint and how they are coupled to maladaptive mechanisms in the peripheral and central nervous system leading to persistent joint pain is still poorly understood. Importantly, animal models of RA have proven useful in understanding disease pathophysiology and the development of disease modifying agents. More frequent utilization of these models in pain research may guide us to new discoveries.

Recent advances in the development of new DMARDs may provide new opportunities also from a pain perspective. For example janus kinase (JAK) inhibitors, like tofacitinib and baricitinib are newly approved DMARDs that were developed with the aid of preclinical research [23,115,185] and reported to improve RA associated pain in patients [42,56,60], even before disease modification in one study [171]. Exploring the role of JAK/STAT signaling in arthritis-induced pain represents an exciting area of research, especially given earlier work pointing to involvement of spinal JAK/STAT in nerve injury induced pain [47]. Another emerging strategy is use of bispecific antibodies that target multiple cytokines, with the idea being that targeting a single cytokine may be insufficient due to over lapping functions of multiple cytokines. Bispecific antibodies against IL-17A and IL-17F, IL-17A and TNF, and IL-1β and IL-17 are under development and in clinical trials for RA [9]. IL-1β and IL-17 bispecific antibodies decrease joint inflammation, proinflammatory cytokine levels and bone erosion in the CIA model [148,178,182], but the effect on pain has not been investigated. Blocking the effect of multiple cytokines may provide additative or synergistic effect in terms of pain relief, especially if targeting factors that are key components of the innate and adaptive immune system at the same time.

Current guidelines recommend using non-steroidal anti-inflammatory drugs (NSAIDs) to manage RA pain. Studies utilizing the CAIA [10], K/BxN [29,134], AIA (refs) and CIA [73] models show that NSAIDs and inhibition of certain cytokines decrease pain-like behavior only during active disease. However, both preclinical RA models and clinical research have suggested a neuropathic component to RA-associated pain. Thus, the use of drugs that are currently used to treat neuropathic pain may be useful to treat RA-associated pain in the absence of inflammation. Use of RA models with distinct characteristics can likely help elucidate when one treatment may be beneficial over another.

In addition to pharmacological strategies, the European League Against Rheumatism (EULAR) guidelines for RA pain management with non-pharmacological interventions were recently published. They recommended physical activity/exercise regimes and diet/weight management to help manage RA pain [59]. The effects of physical activity and exercise are increasingly being characterized in a number of pain models [62,136,146,154], but not yet RA-associated pain. Similarly, diet modulation of pain is a growing preclinical research area, partially due to the effect of diet on inflammation. High fat diets increase mechanical and thermal hypersensitivity in naïve mice [173], and in mice receiving paw [173], or knee intra-articular knee CFA injections [103]. CIA mice fed a high-fat diet have increased signs of inflammation and an earlier arthritis onset [81,168], but the effect on pain was not been investigated. Thus, diet and exercise modulation of pain in preclinical RA models are areas ready for reverse translational research to better understand the mechanisms behind these interventions.

In conclusion, pain relief following DMARD introduction is in part most likely attributable to suppression of joint inflammation. Still, pain remains problematic in subpopulations of RA patients. Several of the animal models of RA described in this review recapitulate this feature, thus they have the potential to aid in advancing our understanding of mechanisms that underpin the "remaining pain". As the complexity of RA pathology is being unraveled, exploring the role of factors that contribute to RA pathology, but have not yet been investigated from a pain perspective, could open doors to more effective pain-relieving strategies. Based on current data generated with RA models in pain studies, one may predict that targeting certain factors, pathways or mechanisms will be more effective at controlling pain than targeting others. Moving forward, more studies should include male and female rodents to avoid potential confounding factors and possibly identify exciting sex-dependent differences. Continued use of preclinical RA models based on bedside-to-bench principals will ultimately continue to delineate complex pain mechanisms and aid the development and deployment of novel and existing therapies, not only for pain in RA but also pain in other conditions.

**Acknowledgements:** This work was supported by European Union's Seventh Framework Programme (FP7/2007 - 2013) under grant agreement No. 602919 (GLORIA), European Union's Horizon 2020 research and innovation program under the Marie Skłodowska-Curie grant agreement No. 642720 (BonePain), Swedish Research Council (2013-8373), Knut and Alice Wallenberg Foundation (CIS), William K Bowes Foundation (CIS) and IASP John J. Bonica Trainee Fellowship (EK). The authors have no conflicts of interest.

**Table 1: Model overview**

|  | CIA | AIA | SCW | CAIA | K/BxN | ACPA | TNF-tg |
|---|---|---|---|---|---|---|---|
| **Trigger** | CII/ adjuvant | mBSA/ adjuvant | Bacterial cell wall | Anti-CII Ab | Serum/ Anti-GPI Ab | ACPA | hTNF over-expression |
| **Poly/mono** | Poly | Mono | Mono | Poly | Poly | Poly | Poly |
| **Similarities to Human RA** | Adaptive immune system activation against endogenous joint epitopes<br><br>Inflm<br><br>Immune cell infiltration<br><br>Joint destruction | Adaptive immune system activation against exogenous epitope<br><br>Inflm<br><br>Immune cell infiltration<br><br>Joint destruction | Inflm<br><br>Immune cell infiltration<br><br>-Joint destruction<br><br>Recurrent flares | Joint specific ab<br><br>Inflm<br><br>Immune cell infiltration<br><br>Joint destruction<br><br>Immune cell infiltration<br><br>Three phases of pain | Inflm<br><br>Immune cell infiltration<br><br>Joint destruction<br><br>Three phases of pain | Ab mediated processes<br><br>Bone erosion | Inflm<br><br>Immune cell infiltration<br><br>Joint destruction |
| **Pain Behavior** | Mech<br>Spont<br>Loco | Mech<br>Thermal<br>Gait<br>Guarding | Mech | Mech<br>Therm<br>Loco | Mech<br>Therm<br>Loco | Mech<br>Therm<br>Loco | Mech<br>Therm<br>Tail flick |

Abbreviations: Poly, poly arthritis; mono, monoarthritis; inflm, joint inflammation; ab, antibody; mech, mechanical hypersensitivity; therm, thermal hypersensitivity; loco, decreased locomotive activity

**Table 2: Sex Distribution by Model in Preclinical RA Pain Studies**

| Model | # of Studies | Male | Female | Male and female | Unspecified |
|---|---|---|---|---|---|
| **CIA** [7,12,37,62,73,81,139,140,142] | 10 | 5 | 4 | 1 | 0 |
| **AIA** [19–21,54–57,84,85,97,121,148,149,154,160,164–166,188,202] | 20 | 6 | 10 | 3 | 1 |
| **SCW** [30] | 1 | 0 | 1 | 0 | 0 |
| **CAIA** [3,14,15,114,145,161,170,177] | 8 | 6 | 0 | 2 | 0 |
| **K/BxN** [22,34,36,76,77,104,116,145,178] | 9 | 7 | 0 | 1 | 1 |
| **ACPA** [193] | 1 | 1 | 0 | 0 | 0 |
| **TNF-tg** [74] | 1 | 0 | 0 | 0 | 1 |
| **CIA/AIA** [9,11] | 1 | 0 | 0 | 0 | 1 |
| Total | 51 | 25 | 15 | 7 | 4 |
| **% total** | **100%** | **49%** | **30%** | **13 %** | **8%** |

**Table 3: Factors Implicated in Peripheral and Central Pain Mechanisms**

|  | **CIA** | **CAIA** | **AIA** | **K/BxN** | **TNF-tg** | **ACPA** | **SCW** |
|---|---|---|---|---|---|---|---|
| **Peripheral Factors** | TNF<br>IL-6<br>NGF<br>SOC<br>Antibodies<br>Nerve-injury | TNF<br>IL-6<br>α2δ1<br>Antibodies<br>Nerve-injury<br>PGs<br>ASIC3 | TNF<br>IL-6<br>IL-17<br>CXCL1<br>CXCL5<br>Antibodies | TNF<br>α2δ1<br>TRPV1<br>Antibodies<br>Nerve-injury<br>PGs | TNF | CXCL1<br>CXCL2<br>Antibodies | Antibodies<br>TNF |
| **Central Factors** | IL-1β | TNF<br>TLR4 | TNF<br>IL-17? | TLR4 | TNF | TBD | TBD |

TBD, to be determined